\documentclass[10pt,aps,prl,twocolumn,superscriptaddress,nofootinbib,preprintnumbers]{revtex4-1}

\usepackage[dvipdfmx]{graphicx}
\usepackage{bm}
\usepackage{epsfig}
\usepackage{amsmath}
\usepackage{latexsym}
\usepackage{epstopdf}
\usepackage{bbold}
\usepackage{color}

\newcommand{\beq}{\begin{equation}}
\newcommand{\eeq}{\end{equation}}
\newcommand{\bea}{\begin{eqnarray}}
\newcommand{\eea}{\end{eqnarray}}

\def\bml{{\rm B-L}}

\begin{document}
%%%%%%%%%%%%%%%%%%%%%%%%%%%%%%
%%%%%%%%%% Title, etc. %%%%%%%%%%%
%%%%%%%%%%%%%%%%%%%%%%%%%%%%%%
\preprint{IPMU14-0341}
\title{Does asymmetric dark matter always lead to an anti-neutrino signal?}
\author{Hajime Fukuda, Shigeki Matsumoto and Satyanarayan Mukhopadhyay}
\vspace{0.4cm}
\affiliation{Kavli IPMU (WPI), The University of Tokyo, Kashiwa, Chiba 277-8583, Japan}

%%%%%%%%%%%%%%%%%%%%%%%%%%%
%%%%%%%%%%% Abs. %%%%%%%%%%%
%%%%%%%%%%%%%%%%%%%%%%%%%%%
\begin{abstract}
Under rather generic assumptions, we show that in the asymmetric dark matter (ADM) scenario, the sign of the $\bml$ asymmetry stored in the dark matter sector and the standard model sector are always the same. One particularly striking consequence of this result is that, when the dark matter decays or annihilates in the present universe, the resulting final state always involves an anti-neutrino. As a concrete example of this, we construct a composite ADM model and explore the feasibility of detecting such an anti-neutrino signal in atmospheric neutrino detectors.
\end{abstract}
\maketitle

%%%%%%%%%%%%%%%%%%%%%
%%%%% Introduction %%%%%
%%%%%%%%%%%%%%%%%%%%%
\section{Introduction}

\noindent

The experimental probes of particle dark matter at present are primarily motivated by the most widely studied paradigm of weakly interacting massive particles (WIMP). The appeal of the WIMP hypothesis lies in the fact that a particle with mass typically in the range of 100 to 1000\,GeV correctly leads to the observed dark matter (DM) density $-$ a fact that is largely independent of the details of the model under consideration, assuming a standard thermal history of the universe\,\cite{WIMP}. Intriguing alternatives, however, do exist in the DM model space, which lead to equally generic predictions as in the WIMP scenarios. Asymmetric dark matter (ADM) is one such model which has been widely studied. An ADM with a mass in the range of 1 to 10\,GeV leads to the observed ratio of baryon to DM densities, which is again largely independent of the model details in a standard thermal history of the universe\,\cite{ADM}.

With the above alternative solution to the DM puzzle in mind, it is important to look for experimental signatures that can discriminate an ADM from a WIMP. In order to maintain thermal equilibrium in the early universe, WIMP's must interact with the standard model (SM) particles. In most cases, this leads to unsuppressed WIMP annihilations to SM final states in the present universe as well, which are being looked for in various indirect detection experiments\,\cite{IDD_WIMP}. Usually, for a WIMP particle which is identical to its anti-particle, the annihilation final states include equal number of SM particles and anti-particles of a given variety. Although ADM's must also interact with the SM sector to be in chemical equilibrium in the early universe, there is a crucial difference in the expected decay or annihilation products at present universe. The ADM particles are charged under the $U(1)_\bml$ symmetry, which is conserved in all interactions. This leads to a final state with a non-zero $\bml$ charge, with different numbers of SM particles and anti-particles of a given species $-$ a fact that can be utilized to distinguish an ADM from a WIMP\,\cite{IDD_ADM}.

This has interesting consequences in the signatures of the ADM particle. For example, the same interaction leading to the chemical equilibrium can cause decay or annihilation of the ADM particles at present universe. In such a case, we find that it would always produce final states with a positive $\bml$ charge.\footnote{The convention followed by us fixes the baryon number of the universe observed today (in visible matter) to be positive.} This fact implies that the resulting final states always involve an anti-neutrino. This is because, in the SM, neutrinos are the only stable particles carrying a non-zero $\bml$ charge and are electrically neutral. On the other hand, other stable particles carrying a $\bml$ charge, namely the electron and the proton, are electrically charged and must be produced in pairs to make the final state electrically neutral. Such pairs, however, do not carry a net $\bml$ charge. This fact leads to a rather striking signal in large volume atmospheric neutrino detectors which can separate neutrinos from anti-neutrinos to a good accuracy: an anti-neutrino signal with an energy of 1 to 10\,GeV. We emphasize here that although the prediction for {\em either} only a neutrino {\em or} only an anti-neutrino signal depending upon the $\bml$ charge of the surviving ADM particle was made in past studies, our results show that chemical equilibrium determines the surviving ADM (anti-)particle to always have a positive $\bml$ charge, independent of the interaction term involved.

In what follows, we first prove that the sign of the $\bml$ asymmetry stored in the dark matter sector and the SM sector are always the same under rather generic assumptions on the ADM scenario. Thereafter, as a concrete example, we construct an explicit model of composite ADM, which is a hidden baryon of a confining $SU(3)_{\rm D}$ gauge symmetry. Finally, we discuss its decay signatures in the form of anti-neutrinos and their detectability in ongoing and near future experiments.

%%%%%%%%%%%%%%%%%%%%%%%%%%%%
%%%%%%%%%%% Proof %%%%%%%%%%%
%%%%%%%%%%%%%%%%%%%%%%%%%%%%
\section{B $-$ L asymmetry stored in DM sector}

\noindent
In the ADM scenario, the $\bml$ asymmetry is postulated to be generated by a baryogenesis mechanism in the early universe. Leptogenesis\,\cite{Leptogenesis} is one promising way to generate the baryon asymmetry, which at the same time enables us to explain tiny neutrino masses via the see-saw mechanism\,\cite{Seesaw}. Since the ADM particle is charged under $\bml$, the net $\bml$ asymmetry will get distributed between the DM and the SM sectors if there exists an interaction maintaining chemical equilibrium between them. The interaction has to be active after baryogenesis and conserve $\bml$ (but violate the DM number) in order not to wash out the generated asymmetry. The relative magnitudes of the asymmetry stored in each sector can be computed using the detailed balance conditions of the reactions in equilibrium\,\cite{Harvey:1990qw}. In fact, using a general method described in Ref.\,\cite{Weinberg:2008zzc}, we can obtain these relative magnitudes without referring to the details of a specific set of reactions, once the conserved quantum numbers of all the particles in equilibrium are known.

We start with a set of particle species $i$ in the reactions, each having a set of conserved quantum numbers $q_{ia}$, where the index $a$ runs over all the conserved quantum numbers for e.g., $\bml$, hypercharge $Y$, the third component of the weak isospin $T_3$, etc. When the interactions between all the particles are sufficiently weak, the chemical potentials of species $i$ (denoted by $\mu_i$) can be expressed as linear combinations of the chemical potentials of the conserved charges (denoted by $\xi_a$):
\begin{equation}
\mu_i = \sum_{a} q_{ia} \xi_a\,.
\label{eq1}
\end{equation}
In the temperature ($T$) of our interest, all the particles can be considered as highly relativistic. When the chemical potential $\mu_i$ is sufficiently smaller than $T$, the difference between the particle and anti-particle number densities of species $i$ is given by $n_i - \bar{n}_i = \tilde{g}_i \mu_i T^2/6$, with $\tilde{g}_i$ being the spin degrees of freedom, but with an additional factor of $2$ for bosons. With the asymmetries of the conserved charges $A_a \equiv \sum_i q_{ia} (n_i - \bar{n}_i)$, the difference is eventually expressed by the asymmetries:
\begin{equation}
n_i - \bar{n}_i = \sum_{a, b} \tilde{g}_i q_{ia} M^{-1}_{ab} A_b\,,
\,\,\,\,\,
M_{ab} \equiv \sum_i \tilde{g}_i q_{ia} q_{ib}\,,
\label{eq2}
\end{equation}
where $a$ and $b$ run over all the conserved charges, while $i$ runs over all the particles in chemical equilibrium. It is worth noting that the matrix $M$ has a positive definite determinant, and hence is invertible.

We are now in a position to apply Eq.\,(\ref{eq2}) to the ADM scenario with the following three assumptions:
\begin{itemize}
\item
In the DM sector, there is no particle charged under the SM gauge group.
\item
The DM sector chemically decouples from the SM one at the temperature $T_{\rm dec}$ in the early universe.
\item
Then, the $\bml$ asymmetry stored in the DM (SM) sector gives the DM (baryon) density today.
\end{itemize}

First of all, since the only asymmetry with a non-zero value in the early universe is that of the $\bml$ charge, the above equation is simplified as follows:
\begin{equation}
n_i - \bar{n}_i = \sum_a \tilde{g}_i q_{ia} M^{-1}_{a\,1} A_1\,,
\label{eq3}
\end{equation}
where the index `1' denotes $\bml$. The next step is to determine how the net $\bml$ asymmetry of the universe is distributed between the DM and the SM sectors. Since it can be shown that it is sufficient to consider the $2 \times 2$ block of the matrix $M$ in the $\bml$ and hypercharge $Y$ basis, the asymmetry stored in the SM sector is
\begin{equation}
A_{\rm SM} =   \sum_{i \in {\rm SM}}
\tilde{g}_i [q_{i1} q_{i1} M^{-1}_{11} + q_{i1} q_{i2} M^{-1}_{21}] A_1\,,
\label{eq4}
\end{equation}
where $A_{\rm SM} \equiv \sum_{i \in {\rm SM}} q_{i1} (n_i - \bar{n}_i)$ and the index `2' denotes the hypercharge $Y$. For convenience, let us define a matrix $N_{ab}$, which is similar to $M_{ab}$, with the sum going over only the particles in the SM sector:
\begin{equation}
N_{ab} \equiv \sum_{i \in {\rm SM}} \tilde{g}_i q_{ia} q_{ib}\,.
\label{eq5}
\end{equation}
Since none of the particles in the DM sector carries a non-zero hypercharge, i.e., $q_{i2} = 0$ when $i \in {\rm DM}$, we have $M_{12} = N_{12}$ and $M_{22} = N_{22}$. In addition, we also have the inequality $M_{11} > N_{11} >0$, which leads to ${\rm detM} > {\rm detN}$. Using this in Eq.\,(\ref{eq4}), it very simply follows that
\begin{equation}
\frac{A_{\rm DM}}{A_{\rm SM}} = \frac{{\rm detM} - {\rm detN}}{\rm detN} > 0\,,
 \label{eq6}
\end{equation}
where $A_{\rm DM} \equiv \sum_{i \in {\rm DM}} q_{i1} (n_i - \bar{n}_i) = A_1 - A_{\rm SM}$ is the $\bml$ asymmetry stored in the DM sector. Thus, it is very generally true that the $\bml$ asymmetry stored in the DM and the SM sectors have the same sign.\footnote{Our result is a general extension of that in Ref.\,\cite{Ibe:2011hq}, and is not altered even if other new conserved charges exist as long as $\bml$ is the only one shared by both the DM and the SM sectors.}

This fact implies that the ADM particles which survived upto the current epoch always have a positive $\bml$ charge. As a result, when it decays or annihilates into SM particles at present universe, the resulting final state always involves, at least, an anti-neutrino, as mentioned earlier. The only exception is the decay into a stable atom like deuterium, whose branching fraction is however usually subdominant\,\cite{Detmold:2014qqa}.

%%%%%%%%%%%%%%%%%%%%%%%%%%%%
%%%%%%%%%%% Model %%%%%%%%%%%
%%%%%%%%%%%%%%%%%%%%%%%%%%%%
\section{Example : a composite ADM model}

\noindent
The operator responsible for the chemical equilibrium between the DM and the SM sectors is obtained by integrating out all other fields of the DM sector from the Lagrangian of the model under consideration, except the ADM fields. The particular form of the interaction depends on the model details, however, it can always be written as a higher dimensional operator suppressed by some high scale, which is denoted by $\Lambda_{\rm ADM}$ in this letter. Therefore, interactions inducing the decay of the ADM particle becomes particularly interesting when $\Lambda_{\rm ADM}$ is much larger than the electroweak scale.

As an attractive possibility, we consider a composite ADM model, in which the DM sector is described by an unbroken strongly coupled ${\rm SU}(3)_{\rm D}$ gauge theory with $N_f$ vector-like `dark' quarks in the fundamental representation. The ADM particle is then a dark baryon composed of three dark quarks. A composite model allows us to consider the decay of the ADM particle without introducing a very small coupling constant. Furthermore, it predicts a large annihilation cross section of the dark matter and anti-dark matter particles into dark mesons. Such a large annihilation cross section is mandatory in the ADM scenario to eliminate the symmetric component of the DM density from the universe.

Interactions between the DM and the SM sectors are given by higher dimensional operators suppressed by the scale $\Lambda_{\rm ADM}$, which respect the $U(1)_\bml$ symmetry. Among these, the lowest dimension interaction violating the DM number is given by
\begin{equation}
\mathcal{L}_{\rm ADM} = \frac{1}{\Lambda_{\rm ADM}^3}
\left( \bar{\psi}^c_i \psi_j \right)
\left( \bar{\psi}^c_k L \right) H + h.c.\,,
\label{eq: ADM interaction} 
\end{equation}
where $\psi_i$ denotes the quarks in the dark sector, with $i$ being their flavor index, while $L$ and $H$ denote the SM lepton and Higgs doublets respectively. The ${\rm SU}(3)_{\rm D}$ and SU$(2)_{\rm L}$ indices are contracted by the respective epsilon tensors, which are suppressed for brevity. We need at least two flavors in the dark sector, i.e., $N_f \geq 2$, since otherwise the operator vanishes identically.

\begin{figure}[t]
\centering
\includegraphics[width=0.4\textwidth]{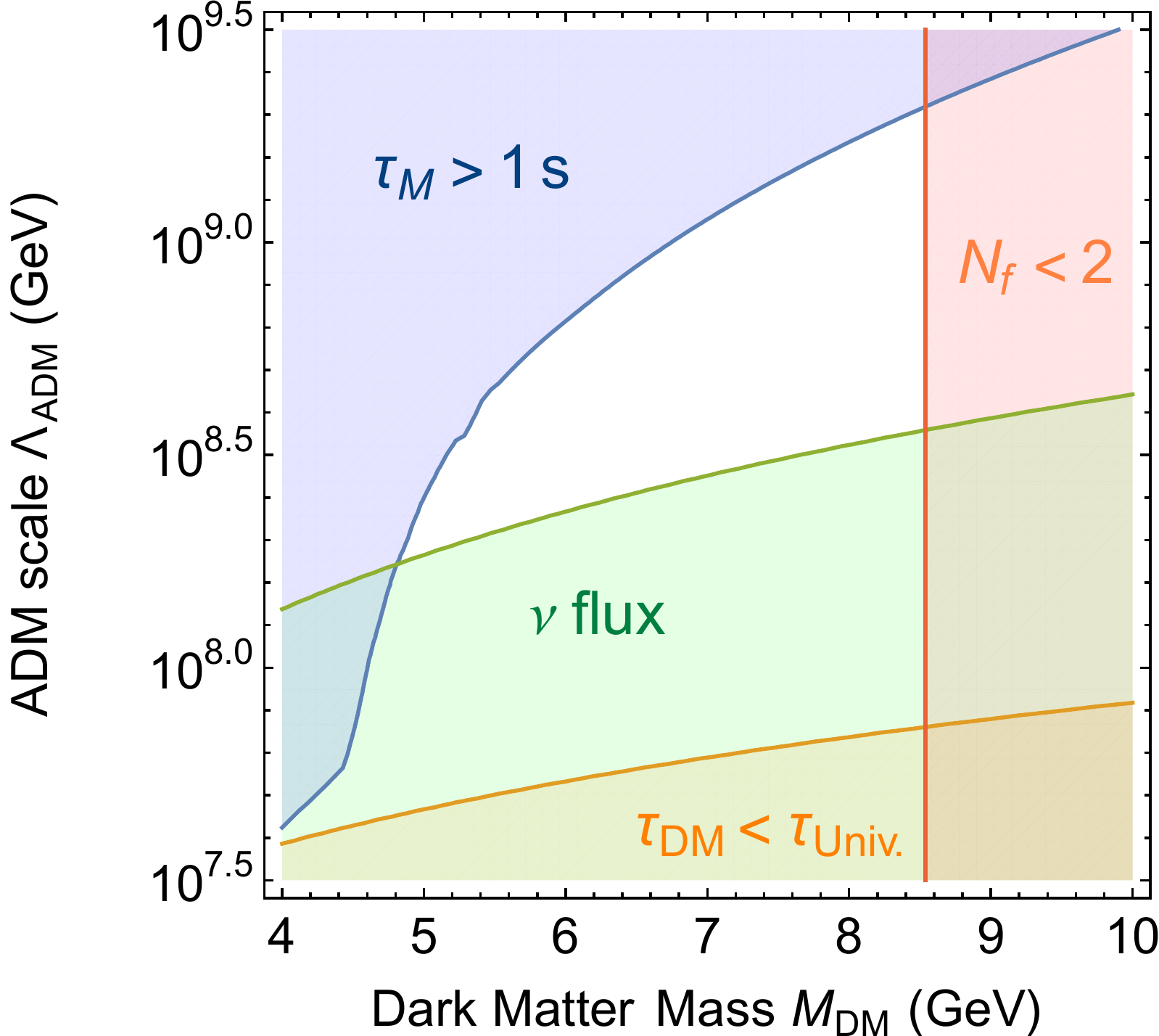}
\caption{\small \sl \label{fig:1} Current constraints on the composite ADM model in the $(m_{\rm DM}, \Lambda_{\rm ADM}$)-plane. See text for details.}
\label{FIGURE}
\end{figure}

This interaction plays the dominant role in maintaining the chemical equilibrium between the DM and the SM sectors, and hence can be used to compute the mass of the ADM particle with the help of Eq.\,(\ref{eq6}). Conservation of $\bml$ then requires the charge of the ADM constituent quarks to be $q_{\psi, \bml} = 1/3$ ($q_{\bar{\psi}, \bml} = -1/3$), and we obtain ${\rm detM} = 79 + 22N_f/3$ and ${\rm detN} = 79$. Here, we assume that there is no new particle charged under the SM gauge group. Using this, the ratio of the $\bml$ asymmetries between the DM and the SM sectors is computed as $A_{\rm DM}/A_{\rm SM} = 22N_f/237$. Remembering the fact that the ratio between $A_{\rm SM}$ and the baryon asymmetry observed today in visible matter (denoted by $A_{\rm B}$) is given by $A_{\rm SM}/A_{\rm B} = 97/30$\,\cite{Harvey:1990qw} and $A_{\rm DM}$ gives the dark matter density today, the mass of the ADM particle is predicted to be
\begin{equation}
m_{\rm DM} = \frac{\Omega_{\rm DM}}{\Omega_{\rm b}}
\frac{3555}{1067 N_f} m_{\rm N}
\simeq \frac{17}{N_f}\,{\rm GeV}\,,
\end{equation}
where $m_{\rm N} \simeq 940$\,MeV, $\Omega_{\rm DM} \simeq 0.12/h^2$ and $\Omega_{\rm b} \simeq 0.022/h^2$ are the nucleon mass, the dark matter and the baryon abundances, respectively, with $h \simeq 0.67$ being the normalized Hubble constant. The condition $N_f \geq 2$ leads to an upper limit on the mass of the ADM particle, $m_{\rm DM} \leq 8.5$\,GeV, as shown in Fig.\,\ref{FIGURE}.

In order to avoid having a large amount of dark radiation in our universe, we introduce mass terms for the hidden quarks $m_\psi$. Assuming that two of the $N_f$ flavors are sufficiently light, the light dark mesons (Nambu-Goldstone (NG) bosons denoted by $M_i$, $i = 1, 2, 3$) have the mass of $m_M^2 \simeq 2m_\psi \Lambda_c^3/f_M^2$, where $\Lambda_c$ and $f_M$ are the scale of the chiral condensate $-\langle \bar{\psi} \psi \rangle^{1/3}$ and the meson decay constant of $SU(3)_{\rm D}$, respectively. The mesons can decay into SM particles by mixing with the SM Higgs boson after electroweak symmetry breaking through the interaction
\begin{equation}
\mathcal L_{M} = \frac{1}{\Lambda_{\rm ADM}} \bar{\psi}_i
i\gamma_5 \psi_j H^\dagger H\,.
\label{eq: Mixing term}
\end{equation}
Using the soft NG boson theorem, the mixing angle between the SM Higgs boson and a light meson $M$ is estimated to be $\sin\theta \simeq 2 v \Lambda_c^3/(f_M \Lambda_{\rm ADM} m_h^2)$ with $v \simeq 246$\,GeV and $m_h \simeq 125$\,GeV being the vacuum expectation value and the mass of the Higgs field, respectively. Needless to say, this is a very small mixing when $\Lambda_{\rm ADM}$ is much larger than the electroweak scale, and hence the properties of the SM Higgs boson are not modified by any appreciable amount.

The light meson can however undergo a late time decay, and in order not to disturb the successful big-bang nucleosynthesis, its lifetime should be shorter than about 1 sec. Using the mixing angle $\sin \theta$, the lifetime can be computed as $\tau_M = (\sin \theta)^{-2}/\Gamma_h(m_M)$, where $\Gamma_h(m_M)$ is the total decay width of a SM-like Higgs boson with a mass of $m_M$\,\cite{HHG}. Since $\tau_M$ becomes shorter when $m_M$ is larger, we fix the hidden quark mass to be $m_\psi = f_M$ as the most conservative choice, which is the maximum possible value if the hidden quark mass term is to be treated perturbatively. Imposing the constraint $\tau_M \lesssim 1$\,sec, leads to an upper limit on $\Lambda_{\rm ADM}$, as shown in Fig.\,\ref{FIGURE}. Here, we adopt simple scaling laws, $\Lambda_{\rm QCD}/m_N = \Lambda_c/m_{\rm DM}$ and $f_\pi/m_N = f_M/m_{\rm DM}$, to estimate $\Lambda_c$ and $f_M$ with $\Lambda_{\rm QCD} \simeq 242$\,MeV and $f_\pi \simeq 92$\,MeV being the chiral condensate of QCD and the pion decay constant. It is worth pointing out that the limit is weakened when we introduce a new scale $\Lambda_M < \Lambda_{\rm ADM}$ assuming physics at that scale does not break the DM number. This would lead to a somewhat larger $\sin\theta$, and a detectable signal in direct detection experiments\,\cite{Bhattacherjee:2013jca}.

The interaction in Eq.\,(\ref{eq: ADM interaction}) induces the decay of the ADM particle. After the Higgs field acquires a vacuum expectation value, it gives rise to a four-Fermi interaction between three dark quarks and one neutrino, suppressed by $\sqrt{2}\Lambda_{\rm ADM}^3/v$. To estimate the width of the dominant decay mode, ${\rm DM} \to \overline{\nu} + M$, we adopt the analogous calculation of proton decay in grand unified theories as follows:
\begin{equation}
\Gamma_{\rm DM} \simeq \frac{3v^2m_{\rm DM}}{64\pi\Lambda_{\rm ADM}^6}
\left|W_0(0)\right|^2 \left(1 - \frac{m_M^2}{m_{\rm DM}^2}\right)^2 \,,
\end{equation}
where the form factor $W_0(0)$ is obtained from the hadron matrix element, $\langle M |(\bar{\psi}^c_i \psi_j) \psi_k |DM \rangle$, which is evaluated to be  $\sim 0.1\,{\rm GeV}^2\,(m_{\rm DM}/m_N)^2$ by a simple scaling of the lattice results\,\cite{Aoki:2013yxa}. It then turns out that the lifetime of the ADM particle, $\tau_{\rm DM} \equiv 1/\Gamma_{\rm DM}$, is longer than the age of the universe when $\Lambda_{\rm ADM} \gtrsim 10^{7.5-8}$\,GeV.

As expected from the theorem proved in the previous section, the decay of the ADM particle in the above model always produces an anti-neutrino, which gives a monochromatic signal with an energy of $E_{\overline{\nu}} \simeq (m_{\rm DM}^2 - m_{\rm M}^2)/(2m_{\rm DM})$. There are existing bounds on an excess in neutrino flux over the predicted atmospheric flux from the measurements by the Super-Kamiokande (SK) collaboration. We follow the analysis developed in Ref.\,\cite{Covi:2009xn}, where the SK data on up-going muon neutrino events from the whole sky collected during April 1996 and July 2001 were used to obtain an upper bound on the excess flux. This translates to a lower limit on $\Lambda_{\rm ADM}$, which is shown in Fig.\,\ref{FIGURE}.

%%%%%%%%%%%%%%%%%%%%
%%%%%%%%%%%%%%%%%%%%

All the current limits on the composite ADM model are summarized in Fig.\,\ref{FIGURE}. As we can see, the mass of the ADM particle is constrained from both above and below, $5\,{\rm GeV} \lesssim m_{\rm DM} \lesssim 8.5\,{\rm GeV}$, and it implies that the number of flavors in the DM sector, $N_f$, can be either two or three. It is also worth emphasizing that since the scale of Leptogenesis, and hence of $\bml$ generation, is expected to be $\mathcal{O}(10^{10})$\,GeV, there exists an order of magnitude window in $\Lambda_{\rm ADM}$ to distribute the $\bml$ asymmetry between the SM and the dark sectors.

The ultimate test of such a scenario would of course be to establish an anti-neutrino line signal, which would then require detectors that can separate neutrinos from anti-neutrinos to a good accuracy. In the absence of a magnetized detector with a sufficiently large volume, this is difficult to achieve. A qualitative discussion on this subject can be found in Ref.\,\cite{IDD_ADM}, where the capabilities of some present and future detectors were reviewed. It turns out that among the present ones, the MINOS detector, having a magnetic field, might be able to establish a signal if the statistics is sufficient, while in future Hyper Kamiokande, DUSEL or INO hold the promise to distinguish a low energy antineutrino signal.

%%%%%%%%%%%%%%%%%%%%%%%%%%%%%%%
%%%%%%%%%%% Summary %%%%%%%%%%%
%%%%%%%%%%%%%%%%%%%%%%%%%%%%%%%
%\section{Summary}
%
%\noindent
{\bf Summary:} We have proved that the relative sign of the $\bml$ asymmetries stored in the DM and SM sectors are the same, and it leads to a striking consequence in ADM phenomenology: an anti-neutrino is always produced when the ADM particle decays or annihilates. We have constructed a composite ADM model to demonstrate this idea and have shown that the decay of the ADM indeed produces a detectable anti-neutrino signal.

%%%%%%%%%%%%%%%%%%%%%%%%%%%%%%%%%%%%%%
%%%%%%%%%%% Acknowledgement %%%%%%%%%%%
%%%%%%%%%%%%%%%%%%%%%%%%%%%%%%%%%%%%%%
%\vspace{0.5cm}

%\noindent
{\bf Acknowledgments:}
%\vspace{0.1cm}\\
\noindent
This work is supported by the Grant-in-Aid for Scientific research from the Ministry of Education, Science, Sports, and Culture (MEXT), Japan [No.~26104009/26287039 (for S.~Matsumoto)], as well as by the World Premier International Research Center Initiative (WPI), MEXT, Japan. 

%%%%%%%%%%%%%%%%%%%%%%%%%%%%%%%%%
%%%%%%%%%%% References %%%%%%%%%%%
%%%%%%%%%%%%%%%%%%%%%%%%%%%%%%%%%


\begin{thebibliography}{99}

\bibitem{WIMP}
%\cite{Lee:1977ua}
%\bibitem{Lee:1977ua} 
  B.~W.~Lee and S.~Weinberg,
  %``Cosmological Lower Bound on Heavy Neutrino Masses,''
  Phys.\ Rev.\ Lett.\  {\bf 39}, 165 (1977).
  %%CITATION = PRLTA,39,165;%%
  %807 citations counted in INSPIRE as of 01 Nov 2014

\bibitem{ADM}
%\cite{Kaplan:2009ag}
%\bibitem{Kaplan:2009ag} 
  D.~E.~Kaplan, M.~A.~Luty and K.~M.~Zurek,
  %``Asymmetric Dark Matter,''
  Phys.\ Rev.\ D {\bf 79}, 115016 (2009).
  %[arXiv:0901.4117 [hep-ph]].
  %%CITATION = ARXIV:0901.4117;%%
  %282 citations counted in INSPIRE as of 01 Nov 2014
Also see,
%\cite{Nussinov:1985xr}
%\bibitem{Nussinov:1985xr} 
  S.~Nussinov,
  %``Technocosmology: Could A Technibaryon Excess Provide A 'natural' Missing Mass Candidate?,''
  Phys.\ Lett.\ B {\bf 165}, 55 (1985).
  %%CITATION = PHLTA,B165,55;%%
  %245 citations counted in INSPIRE as of 01 Nov 2014

\bibitem{IDD_WIMP}
For a review, see, 
%\cite{Bertone:2004pz}
%\bibitem{Bertone:2004pz}
  G.~Bertone, D.~Hooper and J.~Silk,
  %``Particle dark matter: Evidence, candidates and constraints,''
  Phys.\ Rept.\  {\bf 405} (2005) 279.
  %[hep-ph/0404175].
  %%CITATION = HEP-PH/0404175;%%
  %1843 citations counted in INSPIRE as of 01 Nov 2014

\bibitem{IDD_ADM}
%\cite{Feldstein:2010xe}
%\bibitem{Feldstein:2010xe} 
  B.~Feldstein and A.~L.~Fitzpatrick,
  %``Discovering Asymmetric Dark Matter with Anti-Neutrinos,''
  JCAP {\bf 1009}, 005 (2010);
  %[arXiv:1003.5662 [hep-ph]].
  %%CITATION = ARXIV:1003.5662;%%
  %22 citations counted in INSPIRE as of 01 Nov 2014
%\cite{Zhao:2014nsa}
%\bibitem{Zhao:2014nsa} 
  Y.~Zhao and K.~M.~Zurek,
  %``Indirect Detection Signatures for the Origin of Asymmetric Dark Matter,''
  JHEP {\bf 1407}, 017 (2014).
  %[arXiv:1401.7664 [hep-ph]].
  %%CITATION = ARXIV:1401.7664;%%
  %4 citations counted in INSPIRE as of 04 Nov 2014

\bibitem{Leptogenesis}
%\cite{Fukugita:1986hr}
%\bibitem{Fukugita:1986hr} 
  M.~Fukugita and T.~Yanagida,
  %``Baryogenesis Without Grand Unification,''
  Phys.\ Lett.\ B {\bf 174}, 45 (1986).
  %%CITATION = PHLTA,B174,45;%%
  %2261 citations counted in INSPIRE as of 01 Nov 2014

\bibitem{Seesaw}
  T.~Yanagida, in Proceedings of the Workshop on Unified Theories and Baryon Number in the Universe,
  eds.\ O.~Sawada and A.~Sugamoto (KEK report 79-18, 1979);
  %\bibitem{Ramond:1979py}
  M.~Gell-Mann, P.~Ramond and R.~Slansky, 
  in Sanibel Symposium, Palm Coast, Fla., Feb 1979.
  %(hep-ph/9809459).
  Also see,
  %\bibitem{Minkowski:1977sc}
  P.~Minkowski,
  %``mu --> e gamma at a Rate of One Out of 1-Billion Muon Decays?,''
  Phys.\ Lett.\  B {\bf 67}, 421 (1977).

%\cite{Harvey:1990qw}
\bibitem{Harvey:1990qw} 
  J.~A.~Harvey and M.~S.~Turner,
  %``Cosmological baryon and lepton number in the presence of electroweak fermion number violation,''
  Phys.\ Rev.\ D {\bf 42}, 3344 (1990).
  %%CITATION = PHRVA,D42,3344;%%
  %538 citations counted in INSPIRE as of 01 Nov 2014

%\cite{Weinberg:2008zzc}
\bibitem{Weinberg:2008zzc} 
  S.~Weinberg,
  %``Cosmology,''
  Oxford, UK: Oxford Univ. Pr. (2008) 593 p
  %44 citations counted in INSPIRE as of 01 Nov 2014

%\cite{Ibe:2011hq}
\bibitem{Ibe:2011hq} 
  M.~Ibe, S.~Matsumoto and T.~T.~Yanagida,
  %``The GeV-scale dark matter with B-L asymmetry,''
  Phys.\ Lett.\ B {\bf 708}, 112 (2012).
  %[arXiv:1110.5452 [hep-ph]].
  %%CITATION = ARXIV:1110.5452;%%
  %16 citations counted in INSPIRE as of 01 Nov 2014

%\cite{Detmold:2014qqa}
\bibitem{Detmold:2014qqa} 
Another exception can be a final state that  includes a stable $\bml$ charged particle of the DM sector, as in multi-component dark matter models. See, for example,
  W.~Detmold, M.~McCullough and A.~Pochinsky,
  %``Dark Nuclei I: Cosmology and Indirect Detection,''
  arXiv:1406.2276 [hep-ph].
  %%CITATION = ARXIV:1406.2276;%%
  %14 citations counted in INSPIRE as of 13 Nov 2014

\bibitem{HHG}
%\cite{Gunion:1989we}
%\bibitem{Gunion:1989we} 
  J.~F.~Gunion, H.~E.~Haber, G.~L.~Kane and S.~Dawson,
  %``The Higgs Hunter's Guide,''
  Front.\ Phys.\  {\bf 80}, 1 (2000);
  %%CITATION = FRPHA,80,1;%%
  %313 citations counted in INSPIRE as of 11 Nov 2014
%\cite{Gunion:1992hs}
%\bibitem{Gunion:1992hs} 
  %J.~F.~Gunion, H.~E.~Haber, G.~L.~Kane and S.~Dawson,
  %``Errata for the Higgs hunter's guide,''
  hep-ph/9302272.
  %%CITATION = HEP-PH/9302272;%%
  %136 citations counted in INSPIRE as of 11 Nov 2014

%\cite{Bhattacherjee:2013jca}
\bibitem{Bhattacherjee:2013jca} 
  B.~Bhattacherjee, S.~Matsumoto, S.~Mukhopadhyay and M.~M.~Nojiri,
  %``Phenomenology of light fermionic asymmetric dark matter,''
  JHEP {\bf 1310}, 032 (2013).
  %[arXiv:1306.5878 [hep-ph]].
  %%CITATION = ARXIV:1306.5878;%%
  %6 citations counted in INSPIRE as of 13 Nov 2014

%\cite{Aoki:2013yxa}
\bibitem{Aoki:2013yxa} 
  Y.~Aoki, E.~Shintani and A.~Soni,
  %``Proton decay matrix elements on the lattice,''
  Phys.\ Rev.\ D {\bf 89}, 014505 (2014).
  %[arXiv:1304.7424 [hep-lat]].
  %%CITATION = ARXIV:1304.7424;%%
  %12 citations counted in INSPIRE as of 12 Nov 2014
 

%\cite{Covi:2009xn}
\bibitem{Covi:2009xn} 
  L.~Covi, M.~Grefe, A.~Ibarra and D.~Tran,
  %``Neutrino Signals from Dark Matter Decay,''
  JCAP {\bf 1004}, 017 (2010).
  %[arXiv:0912.3521 [hep-ph]].
  %%CITATION = ARXIV:0912.3521;%%
  %43 citations counted in INSPIRE as of 12 Nov 2014
 We have extrapolated the limit shown in Fig.\,7 of this reference  upto $\mathcal{O}(1)$\,GeV in neutrino energies by using a quartic fit function.

\end{thebibliography}
\end{document}